\documentclass{PoS}
\usepackage[round]{natbib}
\usepackage{journals}
\usepackage{amssymb}
\usepackage{subfigure}
\usepackage{textcomp}
\usepackage{graphicx}

\title{New Global VLBI observations of the gravitational lensing system MG J0414+0534}

\ShortTitle{MG~J0414$$+$$0534: new global VLBI 18 cm observations}

\author{\speaker{F. Volino}$^{1}$\thanks{Member of the International Max Planck Research School (IMPRS) for Astronomy and Astrophysics at the Universities of Bonn and Cologne.} , 
O. Wucknitz$^{1}$, R.W. Porcas$^{2}$, E. Ros$^{3,2}$, \newline J.P. McKean$^{4}$, A. Brunthaler${^2}$, J.A. Mu\~noz$^{3}$, P. Castangia$^{5}$ ,\newline M.A. Garrett$^{4,6,7,}$, C. Henkel$^{2}$, C.M.V. Impellizzeri$^{8}$ and A. Roy$^{2}$\\
\llap{$^1$}Argelander-Institut f$\ddot{u}$r Astronomie; Auf dem H$\ddot{u}$gel 71, 53121, Bonn,Germany\\
\llap{$^2$}Max Planck Institut f$\ddot{u}$r Radioastronomie; Auf dem H$\ddot{u}$gel 69, 53121 Bonn, Germany\\
\llap{$^3$}Departament d'Astronomia i Astrof\`isica, Universitat de Val\`encia; E-46100 Burjassot, Val\`encia, Spain\\
\llap{$^4$}ASTRON; Oude Hoogeveensedijk 4, 7991 PD Dwingeloo, The Netherlands\\ 
\llap{$^5$}Osservatorio Astronomico di Cagliari; Loc. Poggio dei Pini, Strada 54, 09012 Capoterra (Ca), Italy\\
\llap{$^6$}Leiden Observatory, Leiden University; Postbus 9513, 2300 RA Leiden, The Netherlands\\
\llap{$^7$}Centre for Astrophysics and Supercomputing; Swinburne University of Technology, Australia\\
\llap{$^8$}National Radio Astronomy Observatory; 520 Edgemont Road Charlottesville, VA 22903-2475, United States\\
Email: \email{fvolino@astro.uni-bonn.de}, \email{wucknitz@astro.uni-bonn.de},
\email{rporcas@mpifr-bonn.mpg.de}, \email{Eduardo.Ros@uv.es},
\email{mckean@astron.nl}, \email{brunthal@mpifr-bonn.mpg.de},
\email{pcastang@oa-cagliari.inaf.it},\email{garrett@astron.nl},
\email{p220hen@mpifr-bonn.mpg.de},\email{violette@nrao.edu}
\email{jmunoz@uv.es} \email{aroy@mpifr-bonn.mpg.de}}

\abstract{The  gravitational lens system MG J0414+0534 is formed by an elliptical galaxy at redshift z~$\sim0.96$ and a quasar at z~$\sim2.64$. 
The system geometry is typical of lensing by an elliptical galaxy with the QSO close to and inside a fold caustic. It shows 4 images of the background source, and a partial Einstein ring is visible at optical wavelengths.
It was observed with a global-VLBI array at $\lambda$\,18~cm in June 2008. We present here the imaging results and a preliminary
lens model constrained by these observations.}

\FullConference{10th European VLBI Network Symposium and EVN Users Meeting: VLBI and the new generation of radio arrays\\
		September 20-24, 2010\\
		Manchester Uk}

\begin{document}

\section{The gravitational lensing system MG~J0414+0534}
\label{intro}
Predicted by general relativity, gravitational lensing (GL) occurs when light rays from a background source travel through the gravitational potential of a foreground mass distribution. Multiple images 
of the same source are produced, with the observed image configuration (relative positions of the lensed images, fluxes ratios and time delays) carrying information on the mass distribution that is causing the effect.
This phenomenon allows us to study mass distributions on different scales (corresponding to different regimes of lensing) and, since it is a purely gravitational effect, it has also been used as 
a cosmological tool to address questions related to dark matter halo properties, non-luminous satellite galaxies and  the large-scale structure distribution in the Universe. \\

The lens system presented here provides a beautiful example of an observed image configuration due to GL by a foreground elliptical galaxy. It was discovered by the VLA-MIT survey \citep{bennett86}, 
and also later by other lens surveys (JVAS/CLASS) \citep{myers03}. It consists of a main elliptical galaxy lens \citep{schechter93} at redshift z$=0.9584 \pm 0.0002$ \citep{tonry99} and four 
images of a QSO at z$=2.639 \pm 0.002$ \citep{lawrence95}  separated by up to $2''$. Hubble Space Telescope (HST) observations \citep{falco97} show a partial Einstein ring connecting 
the two brightest, merging images and a third image. The optical data also show the presence of a luminous satellite, 'X', more than $1''$ away from the main lens whose light distribution is not consistent
with that of an elliptical galaxy \citep{schechter93}.
VLBI observations \citep{patnaik96, porcas98, ros00, trotter00} have resolved the radio structures of the QSO images. 
All show a resolved core region and two jet-like structures \citep{ros00, trotter00}. The complex geometry seen with high angular resolution radio observations and the differences between the lensed images may
confuse the lensing interpretation, but actually can be explained by magnification gradients and the source structure. On the other hand, given that the observed geometry depends on the properties of the lens mass,
its complexity may be seen as providing detailed information on the mass distribution. This system also offers an interesting opportunity for CDM-substructure studies. None of the models cited above is able 
to reproduce the observed flux ratios between the lensed images, and  \citet{minezaki09} infer a lower limit on the mass of a CDM subhalo which would explain the mis-match between the observed 
and modeled flux ratios, in particular the anomalous A1/A2 flux.  \\
The lensed quasar has also been found to host a powerful water maser \citep{impellizzeri08}. An
accurate lensing model, constructed from high-sensitivity VLBI observations will be needed to interpret
the unlensed properties of the water maser system. This was one of the motivations for the study
presented here.

\section{New global VLBI observations at $\lambda$\,18\,cm}
Utilizing all the information that the observed image configuration provides is not easy, however, as the intrinsic (unlensed) source structure and the lens mass distribution need to be modeled at the same 
time. This can be done using the LensClean method \citep{wucknitz04} which fits for the source structure in a non-parametric way and determines the best lens model based on it.   
The results presented here provide additional detailed structural information on the lensed components, which is crucial in order to utilize the Lens Modeling method mentioned above.
\label{new}
\subsection{Observations and data reduction}
\label{obs}
We observed the system at a wavelength of 18\,cm on 7~June~2008 for 19\,hours with a global VLBI array
%%We observed the system at a wavelength of 18\,cm on 07~June~2008 for 19\,hours with a global VLBI array. The following telescopes were used: Effelsberg (100\,m, Germany), Westerbork (14 telescopes used as tied array: 
%%equivalent to 93\,m; The Netherlands), Jodrell Bank (76\,m, United Kingdom), Onsala (25\,m, Sweden), Medicina (32\,m, Italy), Noto (32\,m, Italy), Torun (32\,m, Poland), Shanghai~(25\,m, China), 
%%Urumqi (25\,m, China), Hartebeesthoek (26\,m, South Africa) [members of the European VLBI network (EVN)] , 9 antennas of the Very Long Baseline Array (VLBA) (10 antennas of 25\,m across the United States; we used 
%%[Saint Croix, Hancock, North Liberty, Fort Davis, Los Alamos, Kitt Peak, Owens Valley, Brewster, Mauna Kea]), the phased Very Large Array (VLA, 130\,m equivalent, New Mexico, US) and the Arecibo 
%%telescope (305\,m, Puerto Rico) [our target has a declination of $5^{ \circ}$ which can be tracked for about 2\,hours by this telescope, very close to the maximum tracking time of 2\,h:46m]. 
%%%%%%
%The 19 hours of allocated time for this experiment were scheduled in this way: 5 hours of EVN, 3 hours of EVN and VLBA, 1.5 hours EVN - VLBA - VLA, 2.5 hours EVN - VLBA - VLA - ARECIBO, 
%for the remaining 7 hours VLBA and VLA. 
comprising the following 21 telescopes:
{\bf [EVN]}~~~Effelsberg (100\,m), Westerbork tied-array (equivalent to 93\,m), Jodrell Bank (76\,m), Onsala (25\,m), Medicina (32\,m), Noto (32\,m),
Torun (32\,m), Urumqi (25\,m), Shanghai (25\,m), Hartebeesthoek (26\,m),
Arecibo\footnote{
our target, at declination $5^{ \circ}$, can be tracked for $\sim$2\,hours at Arecibo,
close to the maximum of 2h:46m
}(305\,m).\\
{\bf [US NRAO]}~~~9 of the 25\,m dishes of the VLBA (Saint Croix, Hancock, North Liberty, Fort Davis, Los Alamos, Kitt Peak, Owens Valley, Brewster, Mauna Kea)
and the VLA tied-array (equivalent to 130\,m).
Data were recorded in dual-polarization mode at 512\,Mbps, distributed over 8\,subbands per polarization, each of 8\,MHz bandwidth. There were 16 frequency points (channels) per subband. The integration time 
was 1\,second.  
%The total size of the correlated raw visibilities is about 28 GB. The data were correlated in three different portions: two outer ones were correlated in one single pass, one for the EVN and one 
%for VLBA and VLA, with no more than 16 stations; a middle one having more than 16 stations which required 3 different passes in order to correlate all the baselines.\\
 %%%e.g. antennas were divided in 3 groups, A B and C , with intra group baselines and autocorrelations done twice -- ok I understand this -- , 
 %%%and inter group baselines done once -- don't fully understand this -- 
%%The flux density of the target source is strong enough ( \citet{katz97}) so that full phase-referencing was not necessary. \\
The flux density of the target source is strong enough \citep{katz97} so that phase-referencing was only needed
to create an initial image prior to self-calibration. \\

%The observing cycle consisted of a one-hour scan, regularly repeated over 
%the all duration of the experiment, organized as follows: 15 minutes cycles of alternating scans between the target and the phase referencing source J0422+0219, 3.85~$^{\circ}$ away; one of the fringe 
%finders once per hour for 4 minutes was also observed. For the 2.5 hours when Arecibo was observing as phase referencing source we used J0412+0438, 1.07~$^{\circ}$ degree away. We used as fringe finders
%and calibrators for polarization J$0409+1217$, J$0319+4130$, J$0555+3948$, J$0530+1331$ and $3$C$286$. \\

The data reduction and analysis were performed using the NRAO Astronomical Image Processing System (\texttt{AIPS}, version 31Dec10). The visibility phases were corrected for changes in parallactic angle, 
and \textit{a priori} amplitude calibration was determined from measured system temperatures and antenna gains.  
%An initial delay and rate calibration was first performed by fringe-fitting data from some ``fringe-finder'' scans.
%After applying these corrections,
Standard fringe-fitting was performed on the two phase calibrators\footnote{J$0422+0219$ and J$0412+0438$, 3.85~$^{\circ}$ and 1.07~$^{\circ}$ 
from the target; the second was used when Arecibo was in.}. Changes of gain as a function of frequency (across all the channels) were corrected 
via bandpass calibration. The spectral response of each telescope was analyzed from the data of the fringe-finders, and the final bandpass correction for each telescope was constructed using only the scan 
with the flattest amplitude and phase response.
%Per each telescope a solution was found for every scan but only the best one was used to correct the data (the bp table has one entry per each telescope)    
Subsequently, the two phase-reference sources  were mapped by performing self-calibration (initially phase-only but finally phase and amplitude). The phase and amplitude corrections 
from the best models for both sources were applied to the target, which was then imaged. The data on the lens system were then phase self-calibrated with solution intervals of 4, 2 and 1 minutes. Throughout the whole process, the data 
were never averaged in frequency or time, in order to preserve a large field of view. This resulted in a rather large number of visibilities.

\subsection{Imaging results}
\label{imag}
The J0414+0534 lens system is $2''$ in size, and therefore wide-field techniques are required for imaging \citep{garrett99}. Instead of mapping this
as a single field with milliarcsecond (mas) resolution (which would result in an image
of quite unmanageable size) we used the \texttt{AIPS} task \texttt{IMAGR}, which provides a wide-field mapping procedure based on a multi-field approach. We mapped 3 sub-fields, centered on images B, C and the mid-point of 
images A1 and A2. \\ 

\begin{figure*}
\centering
\caption{The gravitational lens system MG J0414+0534. {\bf Centre:} CASTLES image.
The 4 lensed images (A1, A2, B and C), the main lens galaxy and the luminous satellite X are visible. {\bf Outer panels:} The radio VLBI maps from the new observations.
In each panel the labels identify different regions of the source. The size of the restoring beam is $7.9 \times 2.7$ mas$^{2}$ in position angle --9.2 deg.
The maps display contours of 1.4~mJy/beam$\times (-1,1,2,4,7,16,32)$.
The rms noise levels in the 4 images are 0.18, 0.18, 0.17, 0.09 mJy/beam for A1, A2, B and C, respectively.}
\subfigure[]{\label{whole}\includegraphics[clip, scale=0.27]{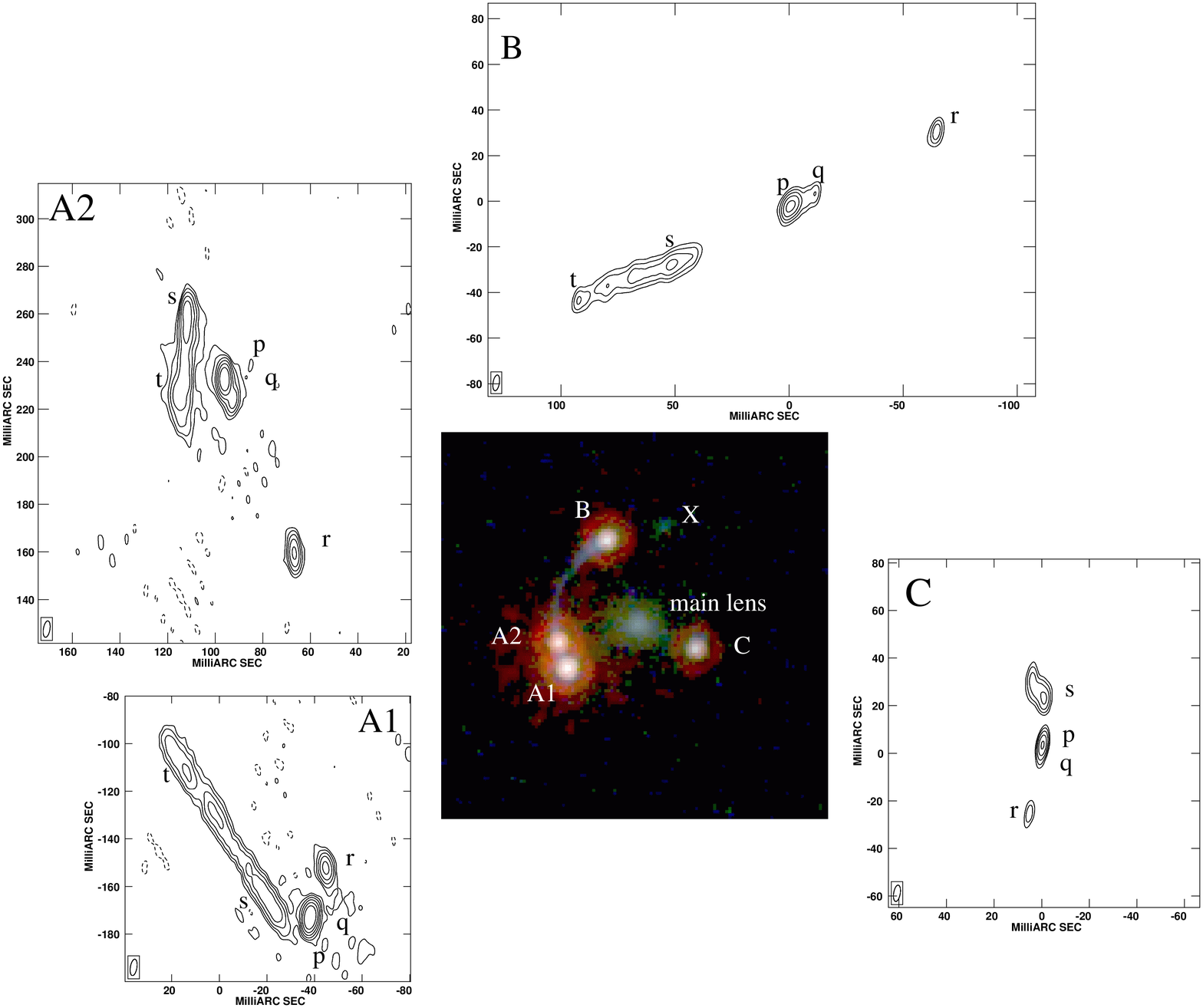}}
\label{maps}
\end{figure*}

Our maps are shown in Fig. \ref{maps}. The central panel is a combination of optical and infrared HST images from the CfA-Arizona Space Telescope LEns Survey (CASTLES) \citep{castle} and the outer panels are the new VLBI maps from the observations described here. A label in each panel identifies the lensed image.  
The restoring beam has a size of \nolinebreak $\simeq 8\times 3~$mas$^{2}$ and position angle (P.A.) $\simeq -9~$deg. 
Previous high resolution VLBI observations of this system \citep{ros00,trotter00} have shown the presence of a resolved core region and 2 jet-like structures in each of the 4 lensed images. 
Our new global VLBI data has higher sensitivity; a comparison of our maps with e.g. Fig. 2 of \citet{trotter00} shows that the structures are consistent with those shown in previous images and, moreover, in images A1, 
A2 and B new components in the outer part of the more elongated jet are clearly seen. We adopt the same notation as these authors to label the different components in the images: 
\textit{p} and \textit{q} for the two core components, \textit{r} for the single jet component on one side of the core, \textit{s} for the more extended jet regions on the other side and \textit{t} the outer part of this. \\
%Existing lens models for this system \citep{ros00,trotter00,minezaki09} predict image B to be the one with the shortest arrival time and, therefore, the least changed by the lensing transformation (??? WHY IS THIS SO ??); this is the image 
%which most resembles the intrinsic (unlensed) source structure which, for example, is highly distorted in the two brightest images A1 and A2.\\

Image B shows a resolved core region extending over $\sim 20$~mas and two 
jet like features (\textit{r} and \textit{s/t}) not exactly aligned with it.
This image is highly stretched in the direction tangential to that of the lensing galaxy. Images A1 and A2 show the most complex structures, highly distorted in the 
tangential directions. Image A1 has the two jet-like features parallel to each other and not aligned with the central core.
The brighter jet (\textit{s/t}) is highly
elongated and extends for more than 80\,mas in the NE direction. In image A2 the two extended jet-like features are also not aligned with the core. 
The brighter (\textit{s/t}) is elongated
in the NS direction over $\sim60$~mas, while \textit{r} is $\sim 80$~mas away in the SW direction.
The core region extends over $\sim 20$~mas, showing a two-component structure. Image C shows a NS structure,
with the core in between \textit{r} and \textit{s/t};
the northern jet component is brighter than the southern, and is more elongated.
This image is the least bright (and hence least resolved) amongst the four.
The total flux densities measured for each subfield agree with the flux ratios between the lensed images given by \citet{katz97}.  
The lens models cited above predict that the two image pairs A1\&A2 and B\&C should each have opposite parities.
The extended structures seen at radio wavelengths clearly confirm this
although the lensing transformations are highly non-linear. 

%\begin{figure*}
%\centering
%\subfigure[]{\label{a2}\includegraphics[width=7cm, angle=360,clip]{A2evn.ps}}
%\centering
%\subfigure[]{\label{b}\includegraphics[width=7cm,clip,angle= 0,clip ]{Bevn.ps}}
%\centering
%\subfigure[]{\label{a1}\includegraphics[width=7cm,angle=0,clip]{A1evn.ps}}
%\centering
%\subfigure[]{\label{c}\includegraphics[width=7cm, angle=0,clip]{Cevn.ps}}
%\caption{Results from global-VLBI 1.8 GHz observations of the system MG J$0414$+$0534$. Each field represents one of the four gravitationally lensed images. The size of the  restoring beam is $7.9 x 2.7$ mas$^{2}$ and 
%its position angle is -9.24 deg. The map display contours of 1.4~mJ/beam$x (-1,1,2,4,7,16,32)$. The root-mean-square (rms) noise level in the 4 images are 0.18,0.17,0.09 mJy/beam for A1 and A2, B and C. 
%\textbf{label the components} }
%\label{maps}
%\end{figure*}
%%%%%%%%
%%%%%%%%

%\begin{figure}
%\label{iband}
%\includegraphics[scale=0.3,clip]{MG0414Iband.eps}
%\caption{i-band image of the system}
%\end{figure}

\section{Lens modeling}
The work done by \citet{ros00} to model this system, using X-band VLBI data as constraints, has shown that the observed image configuration is reproduced quite well by modeling the main lens galaxy as 
a Singular Isothermal Ellipse plus a contribution from external shear, and assuming a Singular Isothermal Sphere for the luminous satellite X.

\citet{trotter00} used C-band VLBA data to represent the gravitational potential as a Taylor expansion.
They find that the mass distribution of the main lens galaxy is highly asymmetric, or a perturber is contributing to the 
lensing effect (see Section 5 in \citealt{trotter00}). 
Both studies find that the luminous satellite X needs to be included in the lens modeling, but they use a rather simple model for the lensed source.

In Table \ref{lresults1} we show  preliminary results from lens modeling which we have initiated, which will be used as a starting point for the more complex 
method mentioned above (Sec. \ref{intro}). As constraints we have used the image positions of the centroid \textit{p} and of feature \textit{r}. \\
  
\begin{table*}
 \centering
%% \begin{minipage}{170mm}
\caption{Best model parameters. The image positions of the core centroid, component \textit{r}, and the optical positions of the main lens and object X were used as constraints.
An accuracy of 0.1\,mas for the relative positions was assumed.
The main lens galaxy was parametrized  as a Singular Isothermal Ellipse with external shear and the satellite X as a Singular
Isothermal Sphere. The lens strengths of the 2 lenses (Einstein radii b$_{ml}$, b$_{x}$), and the ellipticity and shear of the main lens (e, $\gamma$)
were varied while searching for the best values for the ellipse and shear position angles ($\theta_{e}$, $\theta_{\gamma}$). 
Afterwards all the parameters, together with the positions of the lenses,  x$_{ml}$, y$_{ml}$ and x$_{x}$, y$_{x}$ with respect to the core centroid in A1, were varied.}

\vspace{5mm}
  \begin{tabular}{cccccccccc}
  \hline
  \hline 
  b$_{ml}$     &\multicolumn{2}{c}{x$_{ml}$, y$_{ml}$} & e   & $\theta_{e}$ & $\gamma$ & $\theta_{\gamma}$ & b$_{x}$ &\multicolumn{2}{c}{x$_{x}$, y$_{x}$} \\
(mas)           &\multicolumn{2}{c}{(mas)}             &     & (deg)        &          & (deg) & (mas) &\multicolumn{2}{c}{(mas)}\\
  \hline 
  1078.30     &\multicolumn{2}{c}{$-$1070.31, 646.01}  & 0.21& 84.20        &0.08      & $-$55.37           &  149.38  &\multicolumn{2}{c}{$-$1264.46, 1793.64} \\
  \hline
\end{tabular}

%%\end{minipage}
\label{lresults1}
\end{table*}

\section{Summary and Outlook}
Our preliminary results from lens modeling confirm that the main lens galaxy and the luminous satellite X can reproduce the lensed image configuration for the system presented here. 
The values fitted for the ellipticity and sky position angle agree with those constrained by optical data \citep{falco97}. Analysis of these results, and error estimates on the model 
parameters are currently being done. \\ 
%An offset is present between the modeled position of the satellite X and the observed one. 
We plan to refine our results using a non-parametric lens-modeling approach following \citet{wucknitz04}; we will address the question
concerning the radial mass distribution of the main lens galaxy, the position of the satellite X and any other dark substructure which is causing the flux mis-matches.  

\section*{Acknowledgments}
This work was supported by the Emmy-Noether Programme of the Deutsche Forschungsgemeinschaft, reference WU588/1-1. FV was supported for part of this research through a 
stipend from the International Max Planck Research School (IMPRS) for Astronomy and Astrophysics at the Universities of Bonn and Cologne.
%\begin{thebibliography}{99}
%\bibitem{...} 
%....

%\end{thebibliography}

%\end{document}

{
\small
\bibsep0.4ex
\bibliographystyle{aa}
\bibliography{biblio}

\begin{thebibliography}{16}
\expandafter\ifx\csname natexlab\endcsname\relax\def\natexlab#1{#1}\fi

\bibitem[{{Bennett} {et~al.}(1986){Bennett}, {Lawrence}, {Burke}, {Hewitt}, \&
  {Mahoney}}]{bennett86}
{Bennett}, C.~L., {Lawrence}, C.~R., {Burke}, B.~F., {Hewitt}, J.~N., \&
  {Mahoney}, J. 1986, \apjs, 61, 1

\bibitem[{{Falco} {et~al.}(1999){Falco}, {Kochanek}, {Lehar}, {McLeod},
  {Munoz}, {Impey}, {Keeton}, {Peng}, \& {Rix}}]{castle}
{Falco}, E.~E., {Kochanek}, C.~S., {Lehar}, J., {et~al.} 1999, ArXiv
  Astrophysics e-prints

\bibitem[{{Falco} {et~al.}(1997){Falco}, {Lehar}, \& {Shapiro}}]{falco97}
{Falco}, E.~E., {Lehar}, J., \& {Shapiro}, I.~I. 1997, \aj, 113, 540

\bibitem[{{Garrett} {et~al.}(1999){Garrett}, {Porcas}, {Pedlar}, {Muxlow}, \&
  {Garrington}}]{garrett99}
{Garrett}, M.~A., {Porcas}, R.~W., {Pedlar}, A., {Muxlow}, T.~W.~B., \&
  {Garrington}, S.~T. 1999, \nar, 43, 519

\bibitem[{{Impellizzeri} {et~al.}(2008){Impellizzeri}, {McKean}, {Castangia},
  {Roy}, {Henkel}, {Brunthaler}, \& {Wucknitz}}]{impellizzeri08}
{Impellizzeri}, C.~M.~V., {McKean}, J.~P., {Castangia}, P., {et~al.} 2008,
  \nat, 456, 927

\bibitem[{{Katz} {et~al.}(1997){Katz}, {Moore}, \& {Hewitt}}]{katz97}
{Katz}, C.~A., {Moore}, C.~B., \& {Hewitt}, J.~N. 1997, \apj, 475, 512

\bibitem[{{Lawrence} {et~al.}(1995){Lawrence}, {Elston}, {Januzzi}, \&
  {Turner}}]{lawrence95}
{Lawrence}, C.~R., {Elston}, R., {Januzzi}, B.~T., \& {Turner}, E.~L. 1995,
  \aj, 110, 2570

\bibitem[{{Minezaki} {et~al.}(2009){Minezaki}, {Chiba}, {Kashikawa}, {Inoue},
  \& {Kataza}}]{minezaki09}
{Minezaki}, T., {Chiba}, M., {Kashikawa}, N., {Inoue}, K.~T., \& {Kataza}, H.
  2009, \apj, 697, 610

\bibitem[{{Myers} {et~al.}(2003){Myers}, {Jackson}, {Browne}, {de Bruyn},
  {Pearson}, {Readhead}, {Wilkinson}, {Biggs}, {Blandford}, {Fassnacht},
  {Koopmans}, {Marlow}, {McKean}, {Norbury}, {Phillips}, {Rusin}, {Shepherd},
  \& {Sykes}}]{myers03}
{Myers}, S.~T., {Jackson}, N.~J., {Browne}, I.~W.~A., {et~al.} 2003, \mnras,
  341, 1

\bibitem[{{Patnaik} \& {Porcas}(1996)}]{patnaik96}
{Patnaik}, A.~R. \& {Porcas}, R.~W. 1996, in IAU Symposium, Vol. 173,
  Astrophysical Applications of Gravitational Lensing, ed. {C.~S.~Kochanek \&
  J.~N.~Hewitt}, 305

\bibitem[{{Porcas}(1998)}]{porcas98}
{Porcas}, R.~W. 1998, in Astronomical Society of the Pacific Conference Series,
  Vol. 144, IAU Colloq. 164: Radio Emission from Galactic and Extragalactic
  Compact Sources, ed. {J.~A.~Zensus, G.~B.~Taylor, \& J.~M.~Wrobel}, 303

\bibitem[{{Ros} {et~al.}(2000){Ros}, {Guirado}, {Marcaide}, {P{\'e}rez-Torres},
  {Falco}, {Mu{\~n}oz}, {Alberdi}, \& {Lara}}]{ros00}
{Ros}, E., {Guirado}, J.~C., {Marcaide}, J.~M., {et~al.} 2000, \aap, 362, 845

\bibitem[{{Schechter} \& {Moore}(1993)}]{schechter93}
{Schechter}, P.~L. \& {Moore}, C.~B. 1993, \aj, 105, 1

\bibitem[{{Tonry} \& {Kochanek}(1999)}]{tonry99}
{Tonry}, J.~L. \& {Kochanek}, C.~S. 1999, \aj, 117, 2034

\bibitem[{{Trotter} {et~al.}(2000){Trotter}, {Winn}, \& {Hewitt}}]{trotter00}
{Trotter}, C.~S., {Winn}, J.~N., \& {Hewitt}, J.~N. 2000, \apj, 535, 671

\bibitem[{{Wucknitz}(2004)}]{wucknitz04}
{Wucknitz}, O. 2004, \mnras, 349, 1

\end{thebibliography}
}
\end{document}